\documentclass[onecolumn,floatfix,showpacs,aps,nofootinbib]{revtex4}

\usepackage[dvips]{epsfig}
\usepackage[english]{babel}
\usepackage{bbm}
\usepackage{verbatim}
\usepackage{array}
\usepackage{amsmath}
\usepackage{multirow}
\usepackage{amsfonts}
\usepackage{amssymb}
\usepackage{hyperref}
\usepackage{leading}
\usepackage[dvipsnames]{xcolor}
\hypersetup{colorlinks=true,linkbordercolor=Blue,linkcolor=Blue, citecolor=Blue}
\usepackage{subeqnarray}
\usepackage{setspace}
\usepackage{indentfirst} 

\usepackage{gensymb}

\begin{document}

\title{Massive spin one-half one particle states for the mass dimension one fermions}

\author{J. M. Hoff da Silva$^{1}$} \email{julio.hoff@unesp.br}
\author{R. J. Bueno Rogerio$^{2}$} \email{rodolforogerio@unifei.edu.br}

\affiliation{$^{1}$Departamento de F\'isica e Qu\'imica, Universidade
Estadual Paulista, UNESP, Av. Dr. Ariberto Pereira da Cunha, 333, Guaratinguet\'a, SP,
Brazil.}
\affiliation{$^{2}$Instituto de F\'isica e Qu\'imica, Universidade Federal de Itajub\'a - IFQ/UNIFEI, \\
Av. BPS 1303, CEP 37500-903, Itajub\'a - MG, Brazil.}


\begin{abstract}
We study the conditions under which a non-standard Wigner class concerning discrete symmetries may arise for massive spin one-half states. The mass dimension one fermionic states are shown \textcolor{red}{to} constitute explicit examples. We also show how to conciliate these states with the current criticism due to the Lee and Wick, and Weinberg formulation.
\end{abstract}

\pacs{11.30.Er, 03.70.+k}

\maketitle

\section{Introduction}

The seminal work of Wigner in the fall of the thirties has founded the very concept of particle in physics \cite{WIG}. It was shown in a rigid and exhaustive manner that a particle is nothing but an irreducible representation, in the Hilbert space, of the Poincar\'e group. The approach developed in Ref. \cite{WIG}, however, was specifically designed for the orthocronous proper Lorentz subgroup, without taking into account the reflections performed by discrete symmetries that lead to the full Lorentz group. The study of one particle states including reflections was presented by Wigner in Ref. \cite{WIG2}. 

The systematization elaborated by Wigner culminate, after all, in four different cases, among which usual particles all belong to one and the same class. That is, three cases are quite unusual with respect to their responses under action of parity, time reversal, and (an internal symmetry) charge conjugation operations, encoding certain doubling states, i.e., degeneracy, under reflections. Particularly, a specific quantum state holding $(CPT)^2=+1$ for spin one-half representations is reached. 

Revisiting Wigner's results and framing it in a broader scope, Weinberg concludes that the alluded fermionic class found by Wigner does not exist in quantum field theory\footnote{The analysis performed in \cite{WEI} is not restricted to spin one-half particle states, but we shall keep our analysis exclusively to that case.} \cite{WEI}. Moreover, soon after the Wigner's analysis including reflections, Lee and Wick \cite{LW} argued that the quantum field scope of the standard formulation, also taking locality into account, eliminate the possibility of the unusual Wigner classes, including $(CPT)^2=+1$ for any fermionic state. This last fact is also recovered in the continuation of Weinberg's formulation.   

In this paper we shall discuss and contrast this controversy with the possibility of spin one-half particle states endowed of canonical mass dimension one \cite{DHA}. These particles are built to serve, from the realm of quantum field theory, as a dark matter candidate. After the subtleties of the formulation, the resulting particle states may be studied by means of its response under discrete and internal symmetries operations and it turns out that it fills the aforementioned case: $(CPT)^2=+1$. In a previous paper \cite{AHL} it was show a way to circumvent the so-called Weinberg's no go theorem on the (non)existence of spin one-half fields except the usual Dirac one (apart, possibly, from Majorana fields). The whole argumentation of Ref. \cite{AHL} was based on the truly need of a new spinorial dual structure. The argumentation presented in Ref. \cite{AHL} is mathematically well-posed and consistent, nevertheless here we shall approach the controversy at both equivalent levels, particle and field, without taking into account the theory of duals. As we shall see, it is possible to conciliate all the perspectives at both levels by evincing some sharp points in the formulation presented in Ref. \cite{DHA}. At some extent, several pieces leading to our argumentation already exist, but here we concatenate their crucial aspects contrasting them with the standard formulation. Thus, we make explicit some bifurcation points, so to speak, in the quantum representation of the Poincar\'e algebra and in the formulation of the corresponding quantum field, as well. As a common point, we stress the necessity of a sector in the Hilbert space to accommodate one particle states via degenerate eigenvectors of the set $\{\vec{{\bf P}}, H, J_\sigma\}$ (where ${\bf P}$ is the momentum operator, $H$ is the Hamiltonian and $J_\sigma$ stands for a relevant spin operator) with eigenvalues $(0,m,\pm 1/2)$, respectively. These degeneracy shall be lifted by another label, say $h$, whose appreciation may specify a different type of spinor describing the particle and associated quantum field. This remark is presented as a required condition to the alluded Wigner class, but we also complement the analysis showing its complementary condition. 
 
This paper is organized as follows: in the next section we evince the Wigner's result pertinent to our case and contrast it to the hypothesis of existence of specific degenerate states, in the terms previously stated, in the Hilbert space. We then move to the resulting field formulation encompassing the Lee and Wick criticism. Our goal is to show necessary and complementary conditions to the existence of at least one of the non standard Wigner classes, making contact with the field and corresponding particle proposed in Ref. \cite{DHA}.    

\section{Discussion}

We shall begin the discussion by appreciating one particle states. Hereafter we choose all the relevant phases of $C$, $P$ and $T$ in order to achieve an invariant vacuum state\footnote{That is to say, the vacuum state is an eigenstate of the discrete operators with eigenvalue $+1$.}. Concerning spin one-half representations, the results obtained by Wigner in \cite{WIG2} may be summarized as it follows:
\begin{table}[h]
\centering
\begin{tabular}{r|lr}
Class & $(CPT)^2$ & $T^2$ \\ 
\hline                              
1 &  -1  & -1 \\
2 & -1  &  +1 \\
3 &  +1 &  -1 \\
4 &  +1 &  +1 \\
\end{tabular}
\caption{Four classes of spin one-half particles, under reflections, according to Wigner.}
\end{table} 

The main concern regarding Wigner classes $3$ and $4$ arises along the following reasoning: massive spin one half one particle states, say $\Psi_{k,\pm 1/2}$, belonging to the Hilbert space $\mathcal{H}$, are defined as eigenstates of $\{ \vec{{\bf P}}, H, J_\sigma\}$ with eigenvalues $0$, $m$, and $\pm 1/2$ respectively. Here, obviously, $k$ is the rest four momentum. Moreover, denoting by $\mathcal{P}$ the parity matrix acting upon vectors in a direct representation, it is well known that
\begin{eqnarray}
PJ^{\rho\sigma}P^{-1}=\mathcal{P}_\mu^{\;\rho}\mathcal{P}_{\nu}^{\;\sigma}J^{\mu\nu},\\
P{\bf P}^\rho P^{-1}=\mathcal{P}_\mu^\rho{\bf P}^\mu,
\end{eqnarray} 
where $J_{\mu\nu}$ and ${\bf P}_\mu$ are the Lorentz group Rotation/Boost and translations generators, respectively. Hence, one is able to assert that the state $P\Psi_{k,\pm 1/2}$ is also an eigenstate of $\{ \vec{{\bf P}}, H, J_\sigma\}$ with the very same set of eigenvalues. If, and only if, there are no degeneracies, both these states $P\Psi_{k,\pm 1/2}$ and $\Psi_{k,\pm 1/2}$ may differ by a phase at most. The road to the claimed inevitability of fermionic Dirac particles in quantum field theory starts precisely by identifying these states. In particular, being $P\Psi_{k,\pm 1/2}\propto \Psi_{k,\pm 1/2}$ one definitely arrives at $(CPT)^2\Psi_{k,\pm 1/2}=-\Psi_{k,\pm 1/2}$ without any doubt \cite{WEI}. 

We would like to point out that the possible existence of a non empty degenerate sector in the Hilbert space $\mathcal{H}_{D} \subset \mathcal{H}$ may accommodate states whose behavior under reflections belongs to Wigner class 3. Following Wigner's clue on doubling states \cite{WIG2}, this is a necessary condition to the existence of spin one-half realizations of Wigner class 3. All the one particle states studied in Ref. \cite{DHA} belong to $\mathcal{H}_{D}$. 

The action of a discrete symmetry operator in one level, say particle states, determines the action on the other two levels (expansion coefficients and creation and annihilation operators) \cite{TIC}. A direct computation of reflections on the expansion coefficients is built in Ref. \cite{DHA} evincing that such spinors give rise to Wigner class 3 particles. In fact, Ref. \cite{DHA} shows the existence of four different spinors whose associated particles belongs to $\mathcal{H}_{D}$. In this regard $P$, acting on the alluded particle states, perform an endomorphism of $\mathcal{H}_{D}$ on itself.

In order to appreciate the connection between one particle states in $\mathcal{H}_{D}$ and the (criticisms to) Wigner class 3, we move to the point raised in Ref. \cite{LW}, where the authors show that Wigner classes 2, 3, and 4 do not occur (or can be reduced to class 1). The starting point is the Definition 3 (Eq. 2.7) of \cite{LW}. This crucial definition is nothing but the formal account of the transformation of the Dirac field $\psi$ under space inversion
\begin{eqnarray}
P\psi(x)P^{-1}=\chi \bar{P}\psi(\mathcal{P}x), \label{usual}
\end{eqnarray} where $\bar{P}$ stands for the action of the parity operator at the expansion coefficient level and $\chi$ is a relative phase - not relevant for the present discussion. Along with a similar expression for time reversal, it is fairly simple to see, among other things, that $P^2=T^2$ for this standard case. As a result, one is forced to conclude that the only possible Wigner class is indeed class 1, such a \emph{expected} result is precisely achieved because any trace of degenerate state was not taken into account, as one can see in \cite{WEI}.   

On the other hand, endowed with the premise of a non empty $\mathcal{H}_{D}$, let us build up the analog of (\ref{usual}) within this degenerate sector of the Hilbert space. The definition of $P$ action given by     
\begin{eqnarray}
P: &\mathcal{H}_{D}\rightarrow \mathcal{H}_{D}\nonumber\\ &\Psi_{k,\pm}\mapsto P\Psi_{k,\pm}=\eta\Psi '_{k^{'},\pm}, \label{dife}
\end{eqnarray} where $\eta$ is a phase to be further determined as necessary and $k^{'} = (E,-p)$. It is important here to call attention to another relevant aspect of the spinors formal structure found in \cite{DHA}. These spinors are constructed upon an helicity basis with specific phases chosen to ensure to all the four spinors to hold conjugacy under charge conjugation operator. Let us reserve the label $h$ for specifying an element of $\mathcal{H}_{D}$ in such a way that (\ref{dife}) is better written by $ P\Psi_{k,\pm, h}=\Psi '_{k,\pm, h'}$. The $h$ label, differentiating states in $\mathcal{H}_{D}$, shifts the degeneracy among these states but obviously keep the particle species unaffected. As $\Psi_{p,\pm, h}$ is an element of the Hilbert space, it is indeed the case that $\Psi_{p,\pm, h}=a^{\dagger}(\vec{p}, \pm, h)\Psi_{vac}$. Therefore
\begin{equation}
Pa^{\dagger}(\vec{p}, \pm, h)\Psi_{vac}=\eta a^{\dagger}(-\vec{p}, \pm, h')\Psi_{vac}
\end{equation} and remembering that the vacuum state is invariant under $P$ we have 
\begin{equation}
Pa^{\dagger}(\vec{p}, \pm, h)P^{-1}=\eta  a^{\dagger}(-\vec{p}, \pm, h'). \label{rem}
\end{equation} It is important to emphasize the physical content encoded in Eq. (\ref{rem}). As usual, it is still saying that under parity the creation operator at some point $x$ is transformed into a creation operator at $\mathcal{P}x$. Of course, a similar remark may be settled for the annihilation operator. Here we point that the $h$ label must now be also taken into account. Eq. (\ref{rem}) is the creation operator transformation counterpart of (\ref{dife}). The influence of the existence of states in $\mathcal{H}_{D}$, as defined, shall then be inherited by the field operators built with\footnote{In this regard we observe, by passing, that the inhomogeneous transformations to be represented in $\mathcal{H}_{D}$ are attained by the usual semi-simple extension $L_+^\uparrow\Join\mathbb{R}^4$ and, as such, the field decomposition in terms of $a(\vec{p},\pm, h)$ and $a^\dagger(\vec{p},\pm, h)$ is precisely the usual one.} $a(\vec{p},\pm, h)$ and $a^\dagger(\vec{p},\pm, h)$. Here we are focusing in the creation field, mentioning the annihilation field case only when necessary. This short program shall lead us to the analog of Eq. (\ref{usual}) for the states of $\mathcal{H}_{D}$. 

The creation field is given by 
\begin{eqnarray}
\psi^{-}(x)=\frac{1}{(2\pi)^{3/2}}\sum_{h}\int d^3p \; \lambda(\vec{p}, \pm, h)e^{-ip\cdot x}a^{\dagger}(\vec{p}, \pm, h). \label{crea}
\end{eqnarray} Notice here the role played by $h$: usually, when defining the creation (or annihilation field) one takes linear combination in all dependences of $a^\dagger$ (or $a$). It may include the particle species too, but the invariance of interaction terms under Poincar\'e transformations leads to certain constraints such that the species are to be defined as an input, very much like a free label. In the case at hand the spinors entering in the expansion coefficients are built via
spin projections along the momentum \cite{DHA,DS}. Different spinors are endowed with different projections. This is the discrete label over which the linear combination must be carefully taken and hence the sum over $h$. Now, as a matter of fact, there are only two possibilities for $h$ in (\ref{crea}) \cite{DHA}. Let us call these spinors by $\lambda_1(\vec{p}, \pm, h_1)$ and $\lambda_2(\vec{p}, \pm, h_2)$. From Eqs. (\ref{rem}) and (\ref{crea}) it can be readly verified that 
\begin{eqnarray}
P\psi^{-}(x)P^{-1}=\frac{-\eta}{(2\pi)^{3/2}}\int d^3p \Big\{\lambda_1(-\vec{p}, \pm, h_1) a^{\dagger}(\vec{p}, \pm, h_2)&+&\nonumber\\
\lambda_2(-\vec{p}, \pm, h_2) a^{\dagger}(\vec{p}, \pm, h_1)\Big\}e^{-ip\cdot \mathcal{P}x}.\label{indo}
\end{eqnarray} Remembering that $\bar{P}=\bar{P}^{-1}$, with $\bar{P}$ as previously defined, the momentum reversed expansion coefficients are related to the rest momentum coefficients by  
\begin{eqnarray}
\lambda_{i}(-\vec{p}, \pm, h_i)=N \bar{P} \mathcal{D}(L(\vec{p})) \bar{P}\lambda_{i}(\vec{0}, \pm, h_i),\label{mais}
\end{eqnarray} where $i=1,2$, $L(\vec{p})$ is a standard boost responsible to take $k^\mu=(m,\vec{0})$ into $p^\mu$ and $\mathcal{D}(L(\vec{p}))$ its matrix representation. $N$ is a normalization factor not relevant to our purposes. 

At this point we are faced again with the subtle novelty presented by the states belonging to $\mathcal{H}_{D}$: in order to respect the one particle case, or equivalently the definition (\ref{dife}), the introduced expansion coefficients, $\lambda_i$, cannot be eigenspinors of the parity operator. In fact, explicit calculations show that these spinors behave instead as \cite{DHA}
\begin{eqnarray}
\bar{P}\lambda_1(\vec{0},\pm,h_1)=i\lambda_2(\vec{0},\pm,h_2),\nonumber \\
\bar{P}\lambda_2(\vec{0},\pm,h_2)=-i\lambda_1(\vec{0},\pm,h_1).\label{longe}
\end{eqnarray} Notice the relative sign between the phases. Again, it appears explicitly in the computations and bring an important element to our discussion. Had the action of $\bar{P}$ on the expansion coefficients has the same sign, then all the premises to the analysis undertaken in \cite{LW,WEI} would be verified. The different sign in (\ref{longe}), followed by the degeneracy feature, is the complementary condition to Wigner class 3. Indeed, working out Eqs. (\ref{mais}) in the light of (\ref{longe}) and returning to (\ref{indo}) we have
\begin{eqnarray}
P\psi^{-}(x)P^{-1}=\frac{i\eta\bar{P}}{(2\pi)^{3/2}}\int d^3p\, \Big\{\lambda_2(\vec{p}, \pm, h_2) a^{\dagger}(\vec{p}, \pm, h_2)&-&\nonumber\\
\lambda_1(\vec{p}, \pm, h_1) a^{\dagger}(\vec{p}, \pm, h_1)\Big\}e^{-ip\cdot \mathcal{P}x}. \label{oba}
\end{eqnarray} 

The relative sign forbids one to recast the right hand side of (\ref{oba}) as something proportional to $\psi^{-}(x)(\mathcal{P}x)$ (as it is usually reached), the same hold for the annihilation field. As the barrier to achieve the standard form is a relative (not overall) sign in both cases, this cannot be overcome by any choice of phases. Therefore, the analogue of (\ref{usual}) is better expressed by 
\begin{equation}  
P\psi(x)P^{-1}\propto \bar{P}\psi'(\mathcal{P}x), \label{eba}
\end{equation} 
with, as shown, $\psi'\neq\psi$. This is important since demonstrates that the starting point used in Ref. \cite{LW} is not valid for states of $\mathcal{H}_{D}$. Of course, even being the premise used in \cite{LW} not valid in the present case, it does not mean that the conclusions must be necessarily rejected. It turns out, however, that a necessary relation to the claims presented in \cite{LW}, namely $P^2=T^2$ cannot be reached with (\ref{eba}). Particularly, what is obtained in the present case is $P^2=1=-T^2$ and Wigner classe 3 cannot be excluded. In order to appreciate this last point we must to delve into the Weinberg analysis of degenerate states under the action of $T$. 

As dark matter candidates, the spinors used to compose the mass dimension one fermionic field must not carry any gauge charge. Notice that this condition excludes also possible dark matter own gauge interactions, i. e., the interactions coming from gauge symmetries in the dark sector. In other words we are dealing with entirely sterile dark matter candidates. This neutrality is directly achieved by requiring that the action of the charge conjugation operator is such that either, in short, $C\Psi_{h_i}=+\Psi_{h_i}$ (self-dual states) or $C\tilde{\Psi}_{h_i}=-\tilde{\Psi}_{h_i}$ (anti self-dual states), with $i=1,2$. This requirement then leads to the aforementioned four states. We note parenthetically that with respect to our previous discussion about parity, the states $\tilde{\Psi}_{h_i}$ play no new role \footnote{The expansion coefficients leading to these states enter in the annihilation field.} and, as mentioned, our results follows accordingly. 

Suppose an arbitrary linear combination of self dual states given by
\begin{eqnarray}
\Psi = a\Psi_{\vec{p},\pm,h_1}+b \Psi_{\vec{p},\pm,h_2}. \label{t1}
\end{eqnarray} 
There are only two possibilities coming from mass dimension one fermions theory: the coefficients $a$ and $b$ are real numbers and the state $\Psi$ is a self-dual state, or the coefficients are pure imaginary numbers and the (anti self-dual) state is better labeled by $\tilde{\Psi}$. Regarding the particles at hand it can be seen that under the action of $T$ we have the following behavior (please, see \cite{livro} for the explicit action of $T$ in the expansion coefficients)
\begin{eqnarray}
T\Psi_{\vec{p},\pm,h_1} \rightarrow i\tilde{\Psi}_{-\vec{p},\mp,h_2},\nonumber\\
T\Psi_{\vec{p},\pm,h_2} \rightarrow -i\tilde{\Psi}_{-\vec{p},\mp,h_1},\nonumber\\
T\tilde{\Psi}_{\vec{p},\pm,h_1} \rightarrow i\Psi_{-\vec{p},\mp,h_2},\nonumber\\
T\tilde{\Psi}_{\vec{p},\pm,h_2} \rightarrow -i\Psi_{-\vec{p},\mp,h_1}, \label{t2}  
\end{eqnarray} 
where the factor $(-1)^{1/2-\sigma}$ and the possible degenerate phase (see below) were omitted.
 
Let us take $a$ and $b$ both real numbers. The conclusions we shall obtain are also valid to the anti self-dual state as well. Acting with $T$ in the state obtained in Eq. (\ref{t1}), taking into account Weinberg's degeneracy phases possibility \cite{WEI} and the results expressed in (\ref{t2}), we have
\begin{equation}
T\Psi=(-1)^{1/2-\sigma} i (e^{i\phi/2} a \tilde{\Psi}_{-\vec{p},\mp,h_2}-e^{-i\phi/2} b \tilde{\Psi}_{-\vec{p},\mp,h_1}). \label{t3}
\end{equation}
 The phases sign related to $h_1$ and $h_2$ could well be changed without any interference in the argument. It is fairly simple to see that under the action of $C$, the state presented in Eq. (\ref{t3}) reads
\begin{equation}
C(T\Psi)=(-1)^{1/2-\sigma} i (e^{-i\phi/2} a \tilde{\Psi}_{-\vec{p},\mp,h_2}-e^{i\phi/2} b \tilde{\Psi}_{-\vec{p},\mp,h_1}).
\end{equation} 
Notice, however, that the initial $\Psi$ state is self-dual and the action of $T$ must preserve this aspect. Therefore, $e^{i\phi/2}=1$, as it is the only possibility leading to $C(T\Psi)=+(T\Psi)$. Finally, with this unique fixing of phases for the case at hands, a simple algebra exercise provide $T^2\Psi=-\Psi$. As a last remark it can be verified that the particles described in Ref. \cite{DHA} respect $(CPT)^2=+1$.

\section{Concluding Remarks and Outlook}

In this report we present a specific example of one particle states with a degeneracy beyond the spin, fulfilling the $(CPT)^2=+1$ Wigner class for spin 1/2. Although in the earlier literature the existence of special degenerate states have been appreciated \cite{WEI}, here we built a formal example. Up to the best of our knowledge, mass dimension one fermions compose the first concrete example of the aforementioned Wigner class. Within this context such a field may carry features beyond standard model, concerning the associated quantum states transformations under parity and time-reversal. Perhaps variations of the mass dimension one fermions proposal, although certainly difficult, may lead to a new range of possibilities to extend the Standard Model of high energy physics.  

It is important to explicitly remark, stressing once again, that the analysis performed in \cite{WEI} and \cite{LW} are obviously right and are in fact applicable to usual quantum field theory. The crucial point within these seminal analysis is that spin one-half quantum states are understood as belonging to, say $\mathcal{H}_{ND}$, the non degenerate  part of the Hilbert space, in the sense here discussed. Actually, it is fairly reasonable to assert that for usual spin one-half representations the Hilbert space is no other than $\mathcal{H}_{ND}$ and, as a consequence, Wigner non standard classes are ruled out. 

Here we shown two conditions under which Wigner class $3$, for spin one-half, may exist. More rigorously, endowed with the assertion that there exist a non empty $\mathcal{H}_{D}$, it is possible to write the entire spin one-half Hilbert space as $\mathcal{H}=\mathcal{H}_{ND}\oplus \mathcal{H}_{D}$. In this regard, the acting of $P$ in $\mathcal{H}$ states is better expressed in terms of the two algebraic ideals defined by 
\begin{eqnarray}
\mathcal{H}/\mathcal{H}_{ND}=\{\Psi \in \mathcal{H}|P\Psi-\eta\Psi=0 \},\nonumber\\
\mathcal{H}/\mathcal{H}_{D}=\{\Psi, \Psi' \in \mathcal{H}|P\Psi-\eta\Psi'=0, \Psi'\neq\Psi\}.\label{fim}
\end{eqnarray} Besides, the action of parity over the expansion coefficients must include a different relative sign, as in Eqs. (\ref{longe}).

We emphasize that, as explicitly shown in \cite{WEI} at the quantum field level, or in Ref. \cite{LOH} at the expansion coefficients level, the Dirac dynamics shall be respected only in case of states belonging to $\mathcal{H}_{ND}$. Therefore, the quantum states in $\mathcal{H}_{D}$ are necessarily endowed with a different canonical mass dimension. The explicit example brought in \cite{DHA} is shown to have mass dimension one, instead the usual $3/2$ case.   

We shall finalize by given a non rigorous attempt to interpret the label $h$ here used to lift the degeneracy in question. As worked out, the Poincar\'e invariance of the mass dimension one spinor is attained by means of a judicious dual spinor theory \cite{DHA,AHL}. Up to our knowledge, the spinors used to built the theory, the correct dual appreciation apart, may carry symmetries from an eight dimension subalgebra of the Poincar\'e algebra (see \cite{CON} for this formulation). One of the (Casimir) invariants is the usual $m^2$, but the other one is quite complicated and with very difficult physical interpretation\footnote{In this specific framework, the second Casimir is recognized as a generalization of the helicity operator, also called ``lightlike helicity'' (or ``light plane helicity'')\cite{PAT, ALW}.} (see table VII of Ref. \cite{PAT} and \cite{LEE} for a discussion on the role of such quantity in constructing quantum fields). Perhaps a reflex of such quantity in the formal spinor structure is the responsible to the specification of states in $\mathcal{H}_{D}$.

\section{Acknowledgments}
JMHS thanks to CNPq grant N$^{\circ}$. 303561/2018-1 for partial support and RJBR thanks CNPq grant N$^{\circ}$. 155675/2018-4 for the financial support. The authors are grateful to the anonymous Referees for very enlightening comments and criticisms.

\end{document}